\documentclass[journal]{IEEEtran}
\usepackage{cite}

\usepackage{import}

\ifCLASSINFOpdf
\else
\fi
\usepackage{amsmath}
\usepackage{amsmath,graphicx}

\usepackage{import}
\usepackage{booktabs}
\usepackage{tablefootnote}
\usepackage{amsfonts}
\usepackage{amssymb}
\usepackage{caption} 
\usepackage[dvipsnames]{xcolor}
\usepackage{lineno}

\begin{document}
%
\title{Multi-Stream End-to-End Speech Recognition}
%
%
%

\author{Ruizhi~Li,~\IEEEmembership{Student Member,~IEEE,}
        Xiaofei~Wang,~\IEEEmembership{Member,~IEEE,}
        Sri~Harish~Mallidi,~\IEEEmembership{Member,~IEEE,}
        Shinji~Watanabe,~\IEEEmembership{Senior Member,~IEEE,}
        Takaaki~Hori,~\IEEEmembership{Senior Member,~IEEE,}\\
        and~Hynek~Hermansky,~\IEEEmembership{Life Fellow,~IEEE}
\thanks{Ruizhi Li, Xiaofei Wang, Shinji Watanabe, and Hynek Hermansky are with Johns Hopkins University (JHU), USA, e-mail: \{ruizhili, xiaofeiwang, shinjiw, hynek\}@jhu.edu}
\thanks{Sri Harish Mallidi is with Amazon, USA, e-mail: mallidih@amazon.com.}
\thanks{Takaaki Hori is with Mitsubishi Electric Research Laboratories (MERL), USA, e-mail: thori@merl.com.}
\thanks{Manuscript received April 19, 2005; revised August 26, 2015.}}

%
%

\markboth{Journal of \LaTeX\ Class Files,~Vol.~14, No.~8, August~2015}%
{Shell \MakeLowercase{\textit{et al.}}: Bare Demo of IEEEtran.cls for IEEE Journals}
%



\maketitle

\begin{abstract}

Attention-based methods and Connectionist Temporal Classification (CTC) network have been promising research directions for end-to-end (E2E) Automatic Speech Recognition (ASR). 
The joint CTC/Attention model has achieved great success by utilizing both architectures during multi-task training and joint decoding. 
In this work, we present a multi-stream framework based on joint CTC/Attention E2E ASR with parallel streams represented by separate encoders aiming to capture diverse information.
On top of the regular attention networks, the Hierarchical Attention Network (HAN) is introduced to steer the decoder toward the most informative encoders.
A separate CTC network is assigned to each stream to force monotonic alignments. 
Two representative framework have been proposed and discussed, which are Multi-Encoder Multi-Resolution (MEM-Res) framework and Multi-Encoder Multi-Array (MEM-Array) framework, respectively.
In MEM-Res framework, two heterogeneous encoders with different architectures, temporal resolutions and separate CTC networks work in parallel to extract complementary information from same acoustics.
Experiments are conducted on Wall Street Journal (WSJ) and CHiME-4, resulting in relative Word Error Rate (WER) reduction of $18.0-32.1\%$ and the best WER of $3.6\%$ in the WSJ eval92 test set.
The MEM-Array framework aims at improving the far-field ASR robustness using multiple microphone arrays which are activated by separate encoders.
Compared with the best single-array results, the proposed framework has achieved relative WER reduction of $3.7\%$ and $9.7\%$ in AMI and DIRHA multi-array corpora, respectively, which also outperforms conventional fusion strategies.

\end{abstract}


%
\IEEEpeerreviewmaketitle

\section{Introduction}
\label{sec:intro}
%
%
%
%

\IEEEPARstart{R}{ecent} advancements in deep neural networks enabled several practical applications of automatic speech recognition (ASR) technology. 
The main paradigm for an ASR system is the so-called hybrid approach \cite{hinton2012deep}, which involves training a Deep Neural Network (DNN) to predict context dependent phoneme states (or senones) from the acoustic features. 
During inference the predicted senone distributions are provided as inputs to decoder, which combines with lexicon and language model to estimate the word sequence. 
Despite the impressive accuracy of the hybrid system, it requires hand-crafted pronunciation dictionary based on linguistic assumptions, extra training steps to derive context-dependent phonetic
models, and text preprocessing such as tokenization for languages
without explicit word boundaries. 
Consequently, it is quite difficult for non-experts to develop ASR systems for new applications, especially for new languages.

End-to-End (E2E) speech recognition approaches are designed to directly output word or character sequences from the input audio signal. 
This model subsumes several disjoint components in the hybrid ASR model (acoustic model, pronunciation model, language model) into a single neural network. 
As a result, all the components of an E2E model can be trained jointly to optimize a single objective. 
\textcolor{black}{Three dominant end-to-end architectures for ASR are Connectionist Temporal Classification (CTC) \cite{graves2006connectionist,graves2014towards,miao2015eesen},  attention-based encoder decoder \cite{chan2015listen,chorowski2015attention} models and recurrent neural network transducers \cite{graves2012sequence,graves2013speech}. }
\textcolor{black}{While CTC efficiently addresses a sequence-to-sequence problem (speech vectors to word sequence mapping) by avoiding the alignment pre-construction step using dynamic programming, it assumes the conditional independence of label sequence given the input.} 
The attention model does not assume conditional independence of a label sequence resulting in a more flexible model. 
However, attention-based methods encounter difficulty in satisfying the speech-label monotonic property. 
\textcolor{black}{There are previous publications to enhance the monotonic behavior in various ways \cite{Tjandra2017LocalMA, Hou2017GaussianPB, luong-etal-2015-effective, raffel2017online, chiu2017monotonic}. 
These studies are similar in a way that they operate local attention on the windowed encoder outputs to enforce monotonicity.
A joint CTC/Attention framework was proposed in \cite{kim2016joint_icassp2017,hori2017advances,watanabe2017hybrid} with the help of monotonic model, CTC, to alleviate this issue. }
The joint model was shown to provide the state-of-the-art E2E results in several benchmark datasets \cite{watanabe2017hybrid}. 

In this work, we propose a multi-stream architecture within the joint CTC/Attention framework. 
Multi-stream paradigm was successfully used in hybrid ASR \cite{mallidi2018practical, hermansky2013multistream, mallidi2016novel, hermansky2018coding} motivated by observations of multiple parallel processing streams in human speech processing cognitive system. 
For instance, forming streams by band-pass filtering the signal with stream dropout was proposed to deal with noise robustness scenario mimicking human auditory process \cite{mallidi2016novel, mallidi2018practical}. 
However, multi-stream approaches have not been investigated for E2E ASR models. 
This paper is an extension of our prior study \cite{wang2019stream}, which successfully applied the proposed multi-stream concept to multi-array ASR. 
In this work, we present a general formulation to multi-stream framework and two practical E2E applications (MEM-Res and MEM-Array) with additional experiments and discussions.
The framework has the following highlights: 

\begin{enumerate}
\item Multiple Encoders in parallel acting as information streams. 
Two ways of forming the streams have been proposed in this work according to different applications: Parallel encoders with different architectures and temporal resolutions \textcolor{black}{operate} on the same acoustics, which we refer to as Multi-Encoder Multi-Resolution (MEM-Res) model; Parallel input speech from multiple microphone arrays are fed into separate but identical encoders, which we refer to as Multi-Encoder Multi-Array (MEM-Array) model.

\item The Hierarchical Attention Network (HAN) \cite{yang2016hierarchical, hori2017attention, libovicky2017attention} is introduced to dynamically combine knowledge from parallel streams.
\textcolor{black}{While one way of information fusion is to apply one attention mechanism across the outputs of multiple encoder \cite{libovicky2017attention}, 
several studies demonstrated benefits of multiple attention mechanisms \cite{Hayashi2018,chiu2018state,vaswani2017attention, yang2016hierarchical, hori2017attention,libovicky2017attention}.
In \cite{pundak2018deep, kim2018dialog}, secondary attention modules provide a way to incorporate additional contextual information beneficial to the tasks. }
Inspired by the advances in hierarchical attention mechanism in document classification task~\cite{yang2016hierarchical}, multi-modal video description~\cite{hori2017attention} and machine translation~\cite{libovicky2017attention}, we \textcolor{black}{adopt} HAN into our multi-stream model.  
The encoder that carries the most discriminative information for the prediction can dynamically receive a higher weight.  
On top of the per-encoder attention mechanism, stream attention is employed to steer toward the stream, which carries more task-related information.
\item Each encoder is associated with a separate CTC network to guide the frame-wise alignment process for each stream to potentially achieve better performance. 
\end{enumerate}

In MEM-Res model, two parallel encoders with heterogeneous structures are mutually complementary in characterizing the speech signal. 
In E2E ASR, the encoder acts as an acoustic model providing higher-level features for decoding. Bi-directional Long Short-Term Memory (BLSTM) has been widely used due to its ability to model temporal sequences and their long-term dependencies as the encoder architecture; 
Deep Convolutional Neural Network (CNN) was introduced to model spectral local correlations and reduce spectral variations in E2E framework \cite{hori2017advances,zhang2016very}.
The encoder combining CNN with recurrent layers, was suggested to address the limitation of LSTM. 
While temporal subsampling in RNN and max-pooling in CNN aim to reduce the computational complexity and enhance the robustness, it is likely that subsampling technique results in loss of temporal resolution. 
The MEM-Res model proposes to combine RNN-based and CNN-RNN-based networks to form a complementary multi-stream encoder.

In addition to MEM-Res, MEM-Array model is one of the other applications of our multi-stream E2E framework.
Far-field ASR using multiple microphone arrays has become important strategies in the speech community toward a smart speaker scena rio in a meeting room or house environment \cite{carletta2005ami, ravanelli2016realistic, Barker2018}.
Individually, the microphone array is able to bring a substantial performance improvement with algorithms such as beamforming \cite{vincent2017analysis} and masking \cite{wang2016oracle}. 
However, what kind of information can be extracted from each array and how to make multiple arrays work in cooperation are still challenging. 
Time synchronization among arrays is one of the main challenges that most distributed setup face \cite{markovich2015optimal}.
Without any prior knowledge of speaker-array distance or video monitoring, it is difficult to estimate which array carries more reliable information or is less corrupted.  

According to the reports from the CHiME-5 challenge \cite{Barker2018}, which targets the problem of multi-array conversational speech recognition in home environments, the common ways of utilizing multiple arrays in the hybrid ASR system are finding the one with highest Signal-to-Noise/Signal-to-Interference Ratio (SNR/SIR) for decoding \cite{du2018theustc} or fusing the decoding results by voting for the most confident words \cite{kanda2018hitachi}, e.g. ROVER \cite{fiscus1997post}.
Similar to our previous work \cite{wang2017stream}\cite{wang2018stream}, combination using the classifier's posterior probabilities followed by lattice generation has been an alternative approach \cite{misra2003new,xiong2018channel,hermansky2018coding}.
\textcolor{black}{Compared to using the fully decoding results with paths pruning, the combination using the posteriors preserves all the information from the test speech as well as the classifier.}
In terms of the combination strategy, ASR performance monitors have been designed \cite{mallidi2015uncertainty}, resulting in a process of stream confidence generation, guiding the linear fusion of array streams.
While most of the E2E ASR studies engage in single-channel task or multi-channel task from one microphone array \cite{ochiai2017unified,braun_2018,ochiai2017multichannel,kim2017end}, research on multi-array scenario is still unexplored within the E2E framework. 
The MEM-Array model is proposed to solve the aforementioned problem.
The output of each microphone array is modeled by a separate encoder. 
Multiple encoders with the same configuration act as the acoustic models for individual arrays. 
Note that we integrate beamformed signals instead of using all multi-channel signals for the multi-stream framework, which is computationally efficient. This design can make use of the powerful beamforming algorithm as well.

\newcommand{\RNum}[1]{\uppercase\expandafter{\romannumeral #1\relax}}
This paper is organized as follows: 
section \RNum{2} explains the joint CTC/Attention model. 
The description of the proposed multi-stream framework including MEM-Res and MEM-Array is in section \RNum{3}. 
Experiments with results and several analyses for MEM-Res and MEM-Array models are presented in section \RNum{4} and Section \RNum{5}, respectively. 
Finally, in section \RNum{6} the conclusion is derived.

\section{Joint CTC/Attention Mechanism}
\label{ssec:ctcatt}

In this section, we review the joint CTC/attention architecture, which takes advantage of both CTC and attention-based end-to-end ASR approaches during training and decoding.

\subsection{Connectionist Temporal Classification (CTC)}
\label{sssec:ctc}

CTC enforces a monotonic mapping from a $T$-length speech feature sequence, $X=\{\textbf{x}_{t}\in \mathbb{R}^{D}|t = 1,2,...,T\}$, to an $L$-length letter sequence, $C=\{c_{l}\in \mathcal{U}|l = 1,2,...,L\}$. Here $\textbf{x}_{t}$ is a $D$-dimensional acoustic vector at frame $t$, and $c_{l}$ is at position $l$ a letter from $\mathcal{U}$, a set of distinct letters. 

The CTC network introduces a many-to-one function from frame-wise latent variable sequences, $Z=\{z_{t}\in \mathcal{U} \bigcup \text{blank} |t=1,2,...,T\}$, to letter predictions with shorter lengths.
\textcolor{black}{This is a many-to-one function because many CTC paths can respond to the same label sequence by merging repeating characters and removing blank symbols.}
With several conditional independence assumptions, the posterior distribution, $p(C|X)$, is represented as follows:
\begin{equation}
\label{f:ctcloss}
p(C|X)\approx \sum_{Z}\prod_{t} p(z_{t}|X) \triangleq p_\textrm{ctc}(C|X),
\end{equation}
where $p(z_{t}|X)$ is a frame-wise posterior distribution, which is often modeled using BLSTM. 
We also define the CTC objective function $p_\textrm{ctc}(C|X)$.
CTC preserves the benefits that it enforces the monotonic behavior of speech-label alignments, avoids the HMM/GMM construction step and preparation of pronunciation dictionary.




\subsection{Attention-based Encoder-Decoder}
\label{sssec:att}

As one of the most commonly used sequence modeling techniques, the attention-based framework selectively encodes an audio sequence of variable length into a fixed dimension vector representation, which is then consumed by the decoder to produce a distribution over the outputs.  
We can directly estimate the posterior distribution $p(C|X)$ using the chain rule:
\begin{equation}
p(C|X)=\prod_{l=1}^{L}p(c_{l}|c_{1},...,c_{l-1}, X) \triangleq p_\textrm{att}(C|X),
\end{equation}
where $p_\textrm{att}(C|X)$ is defined as the attention-based objective function. 
Typically, a BLSTM-based encoder transforms the speech vectors $X$ into frame-wise hidden vector $\textbf{h}_{t}$
If the encoder subsamples the input by a factor $s$, there will be \textcolor{black}{$\lfloor T/s\rfloor$} time steps in \textcolor{black}{$H=\{\textbf{h}_{1},..., \textbf{h}_{\lfloor T/s\rfloor}\}$}. 
The letter-wise context vector $\textbf{r}_{l}$ is formed as a weighted summation of frame-wise hidden vectors $H$ using content-based attention network \cite{chorowski2015attention}:
\begin{equation}
\label{f:han}
\textcolor{black}{\textbf{r}_{l}={\sum}_{t=1}^{\lfloor T/s\rfloor}a_{lt}\textbf{h}_{t}},
\end{equation}
\begin{equation}
a_{lt} = \textrm{ContentAttention}(\textbf{q}_{l-1}, \textbf{h}_t),  
\end{equation}
where ${a}_{lt}$ is the attention weight, a soft-alignment of $\textbf{h}_{t}$ for output $c_{l}$, and $\textbf{q}_{l-1}$ is the previous decoder state. ContentAttention($\cdot$) is described as follows:
\begin{equation}
e_{lt} = \textbf{g}^\top\textrm{tanh}(\textrm{Lin}(\textbf{q}_{l-1})+\textrm{LinB}( \textbf{h}_t)),  
\end{equation}
\begin{equation}
\textcolor{black}{a_{lt}=\textrm{Softmax}(\{e_{lt}\}^{\lfloor T/s\rfloor}_{t=1})}. 
\end{equation}
$\textbf{g}$ is a learnable vector parameter. \textcolor{black}{$\{e_{lt}\}^{\lfloor T/s\rfloor}_{t=1}$} is a \textcolor{black}{$\lfloor T/s\rfloor$}-dimensional vector. LinB($\cdot$) and Lin($\cdot$) represent the linear transformation with or without bias term, respectively. 


In comparison to CTC, not requiring conditional independence assumptions is one of the advantages of using the attention-based model. 
However, the attention is too flexible to satisfy monotonic alignment constraint in speech recognition tasks.

\subsection{Joint CTC/Attention}
\label{sssec:ctcatt}

The joint CTC/Attention architecture benefits from both CTC and attention-based models since the attention-based encoder-decoder is trained together with CTC within the Multi-Task Learning (MTL) framework. 
The encoder is shared across CTC and attention-based encoders.  
And the objective function to be maximized is a logarithmic linear combination of the CTC and attention objectives, i.e., $p_\textrm{ctc}(C|X)$ and $p_\textrm{att}^{\dagger}(C|X)$:
\begin{equation} 
\label{f:mtl}
\mathcal{L}_\textrm{MTL}=\lambda\log p_\textrm{ctc}(C|X)+(1-\lambda)\log p_\textrm{att}^{\dagger}(C|X),
\end{equation}
where $\lambda$ is a tunable scalar satisfying $0\leq\lambda\leq 1$. $p_\textrm{att}^{\dagger}(C|X)$ is an approximated letter-wise objective where the probability of a prediction is conditioned on previous true labels. 


During inference, the joint CTC/Attention model performs a label-synchronous beam search.
The most probable letter sequence $\hat{C}$ given the speech input $X$ is computed according to
\begin{align}
\label{f:jointdec}
    \hat{C}=\arg\max_{C\in \mathcal{U}^{*}} &\{\lambda \log p_\textrm{ctc}(C|X)+(1-\lambda)\log p_\textrm{att}(C|X) \nonumber \\
    &+\gamma \log p_\textrm{lm}(C)\} 
\end{align}
where external RNN-LM probability $\log p_\textrm{lm}(C)$ is added with a scaling factor $\gamma$. 
For each partial hypothesis $h$ in the beam search, the joint score, the log probability of hypothesized label sequence, can be computed as 
\begin{equation}
    \alpha(h)=\lambda \alpha_\textrm{ctc}(h)+(1-\lambda)\alpha_\textrm{att}(h)+\gamma\alpha_\textrm{lm}(h),
\end{equation}
where the attention decoder score, $\alpha_\textrm{att}(h)$, can be accumulated recursively from hypothesis scores from one step before. 
In terms of CTC score, $\alpha_\textrm{ctc}(h)$, we utilize the CTC prefix probability defined as the cumulative probability of all label sequences that have $h$ as their prefix \cite{graves2008supervised,hori2017joint}.
In this work, we use the look-ahead word-based language model to give the RNN-LM score \cite{hori2018end}, $\alpha_\textrm{lm}(h)$. 
This language model enables us to decode with only a word-based model, rather than using a multi-level LM which uses a character-level LM until the identity of the word is determined.

\section{Proposed Multi-Stream Framework}
\label{ssec:memodel}

\begin{figure}[htb]
  \centering 
  \centerline{\includegraphics[width=7cm]{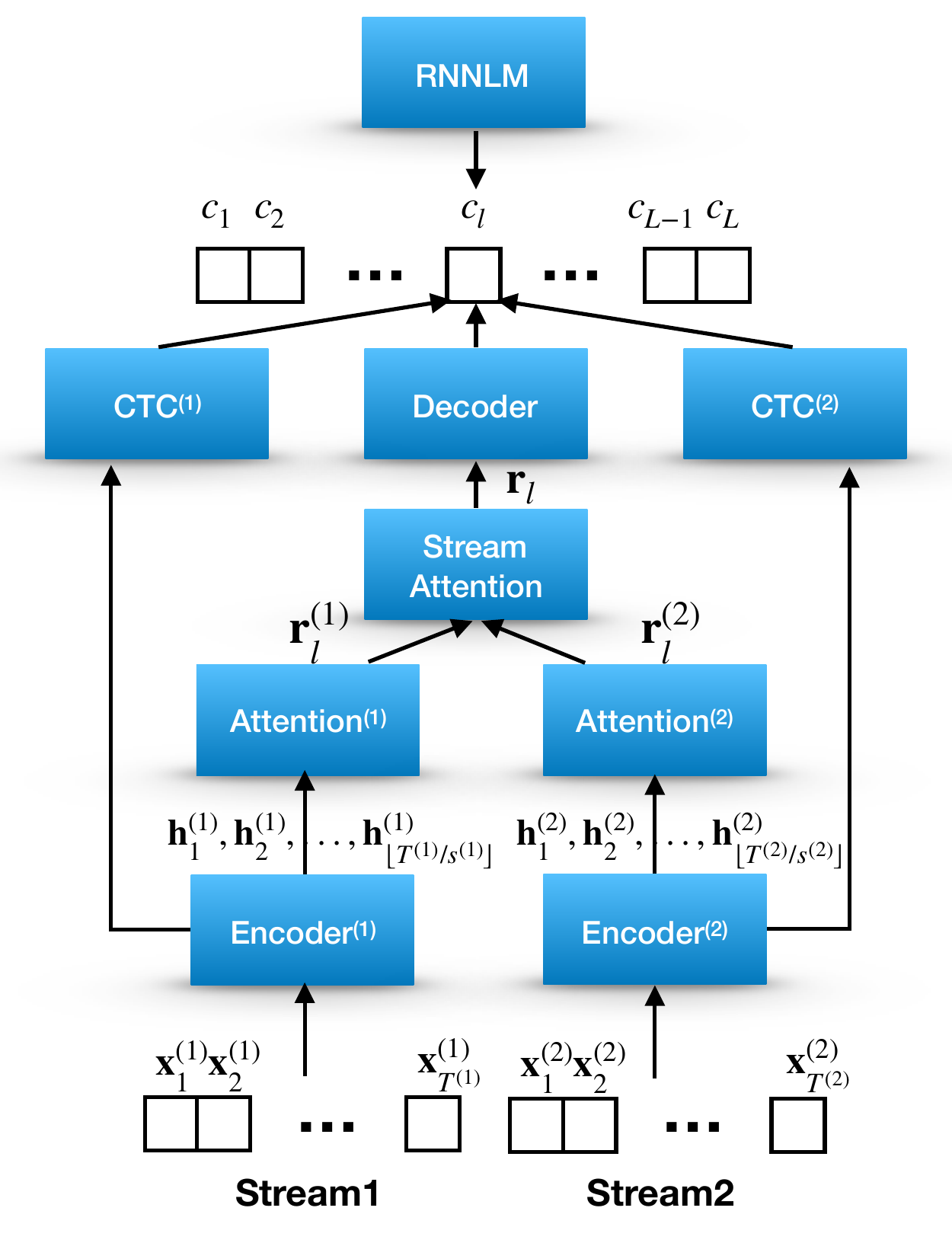}}
  \caption{The Multi-Stream End-to-End Framework.}
  \label{fig:ms}
  \vspace{-0.3cm}
\end{figure}

The proposed multi-stream architecture is shown in Fig. \ref{fig:ms}.
For simplicity to understand the framework, we focus on the two-stream architecture.
Two encoders are presented in parallel to capture information in various ways, followed by an attention fusion mechanism together with per-encoder CTC. 
An external RNN-LM is also involved during the inference step. 
We will describe the details of each component in the following sections. 

\subsection{Parallel Encoders as Multi-Stream}
Similar to acoustic modeling in conventional ASR, the encoder maps the audio features into higher-level feature representations for the use of CTC and attention model:
\begin{equation}
\label{eq:enc}
\textbf{h}_{t}^{(i)}=\textrm{Encoder}^{(i)}(X^{(i)}), i\in\{1, ..., N\}
\end{equation}
where we denote superscript $i\in\{1, ..., N\}$ as the index for ${\textrm{Encoder}^{(i)}}$ corresponding to stream $i$, $\textbf{h}_t^{(i)}$ is the frame-wise hidden vector of stream $i$ introduced in Sec. \ref{sssec:att}.
, and $N$ denotes the number of streams.
$X^{(i)}$ in Eq. \ref{eq:enc} represents a $T^{(i)}$-length speech feature sequence, i.e., $X^{(i)}=\{\textbf{x}_{t}^{(i)}\in \mathbb{R}^{D}|t = 1,2,...,T^{(i)}\}$. Note that it is not mandatory to have frame-level synchronization across all streams since $T^{(i)}, i\in\{1, ..., N\}$, could be different in the proposed model. 
Together with stream-specific subsampling factor $s^{(i)}$, stream $i$ will have \textcolor{black}{$\lfloor T^{(i)}/s^{(i)}\rfloor$} time instances at the encoder-output level. Rounding process of \textcolor{black}{$\lfloor T^{(i)}/s^{(i)}\rfloor$} is performed in the encoder based on different architecture. 

For simplicity, multi-stream model with $N=2$ is depicted in Fig. \ref{fig:ms}, where two encoders in parallel take different input features, $X^{(1)}$ with $T^{(1)}$ frames and $X^{(2)}$ with $T^{(2)}$ frames, respectively.
Each encoder operates on different temporal resolution with subsampling factor $s^{(1)}$ and $s^{(2)}$, where subsampling could be performed in RNN or maxpooling layer in CNN.

\subsection{Hierarchical Attention}
\label{sssec:hieratt}

Since the encoders model the speech signals differently by catching acoustic knowledge in their own ways, encoder-level fusion is suitable to boost the network's ability to retrieve the relevant information.  
We adopt Hierarchical Attention Network (HAN) in \cite{yang2016hierarchical} for information fusion. 
The decoder with HAN is trained to selectively attend to appropriate encoder, based on the context of each prediction in the sentence as well as the higher-level acoustic features from encoders, to achieve a better prediction. 

The letter-wise context vectors, $\textbf{r}_l^{(i)}$, from individual encoders are computed as follows: 
\begin{equation}
\label{f:cv1}
\textcolor{black}{\textbf{r}_{l}^{(i)}={\sum}_{t=1}^{\lfloor T^{(i)}/s^{(i)}\rfloor}a_{lt}^{(i)}\textbf{h}_{t}^{(i)}, i\in\{1, ..., N\}}
\end{equation}
where the attention weights $\{a_{lt}^{(i)}\}$, where \textcolor{black}{$\sum_{t=1}^{\lfloor T^{(i)}/s^{(i)}\rfloor}a_{lt}^{(i)}=1$} , are obtained using a content-based attention mechanism. Note that since encoders perform downsampling, the summations are till \textcolor{black}{$\lfloor T^{(i)}/s^{(i)}\rfloor$} for each individual stream in Eq. (\ref{f:cv1}), respectively.

The fusion context vector $\textbf{r}_l$ is obtained as a convex combination of $\textbf{r}_l^{(i)}, i\in\{1, ..., N\}$, as illustrated in the following: 
\begin{equation}
\label{f:han}
\textbf{r}_{l}={\sum}_{i=1}^{N}\beta_{l}^{(i)}\textbf{r}_{l}^{(i)},
\end{equation}
\begin{equation}
\label{f:l2att}
\beta_{l}^{(i)}=\textrm{ContentAttention}(\textbf{q}_{l-1}, \textbf{r}_l^{(i)}), i\in\{1, ..., N\}.
\end{equation}
The stream-level attention weight, $\beta_{l}^{(i)}$, where $\sum_{i=1}^{N}\beta_{l}^{(i)}=1$, is estimated according to the previous decoder state $\textbf{q}_{l-1}$ and context vector $\textbf{r}_l^{(i)}$ from an individual encoder $i$ as described in Eq. (\ref{f:l2att}).
The fusion context vector is then fed into the decoder to predict the next letter. 

\subsection{Training and Decoding with Per-encoder CTC}
\label{sssec:perctc}
In the CTC/Attention model with a single encoder, the CTC objective serves as an auxiliary task to speed up the procedure of realizing monotonic alignment and providing a sequence-level objective. 
In multi-stream framework, we introduce per-encoder CTC where a separate CTC mechanism is active for each encoder stream during training and decoding.
Sharing one set of CTC among encoders is a soft constraint that limits the potential of diverse encoders to reveal complementary information. 
\textcolor{black}{Sharing CTC refers to the case that linear layers connecting hidden vectors to CTC Softmax layers for each encoders are shared.}
In the case that encoders are with different temporal resolutions and network architectures, per-encoder CTC can further align speech with labels in a monotonic order and customize the sequence modeling of individual streams.

During training and decoding steps, we follow Eq. (\ref{f:mtl}) and (\ref{f:jointdec}) with a change of the CTC objective $\log p_\textrm{ctc}(C|X)$ in the following way:
\begin{equation}
\log p_\textrm{ctc}(C|X)=\frac{1}{N}{\sum}_{i=1}^{N}\log p_{\textrm{ctc}^{(i)}}(C|X),
\end{equation}
where joint CTC loss is the average of per-encoder CTCs. 
In the beam search, the CTC prefix score of hypothesized sequence $h$ is altered as follows: 
\begin{equation}
    \alpha_\textrm{ctc}(h)=\frac{1}{N}{\sum}_{i=1}^{N}\alpha_{\textrm{ctc}^{(i)}}(h),
\end{equation}
where equal weight is assigned to each CTC network. 

\subsection{Multi-Encoder Multi-Resolution}
\label{ssec:MEM-Res}
\begin{figure}[htb]
\vspace{-0.3cm}
  \centering 
  \centerline{\includegraphics[width=7cm]{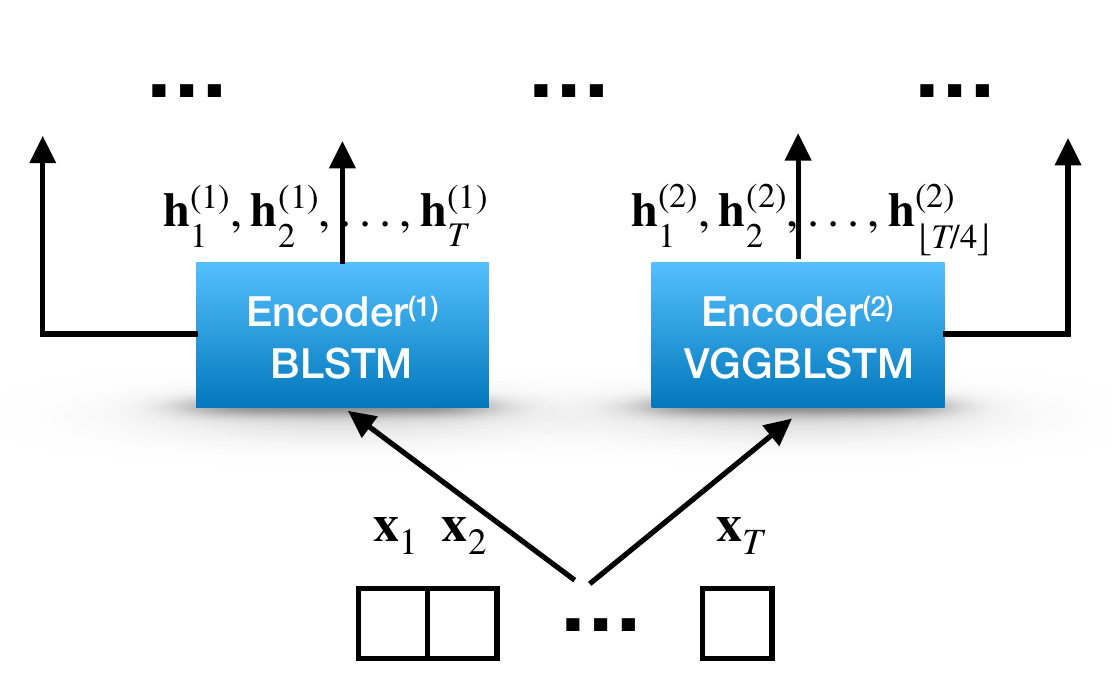}}
\caption{Multi-Encoder Multi-Resolution Architecture.}
\vspace{-0.1cm}
  \label{fig:MEM-Res}
\end{figure}
As one realization of multi-stream framework, we propose a Multi-Encoder Multi-Resolution (MEM-Res) architecture that has two encoders, RNN-based and CNN-RNN-based.
Both encoders take the same input features in parallel operating on different temporal resolutions, aiming to capture complementary information in the speech as depicted in Fig. \ref{fig:MEM-Res}. 

The RNN-based encoder is designed to model temporal sequences with their long-range dependencies. 
Subsampling in BLSTM is often used to decrease the computational cost, but performing subsampling might result in lost information which could be better modeled in RNN. 
In MEM-Res, the BLSTM encoder has only BLSTM layers that extract the frame-wise hidden vector $\textbf{h}_t^{(1)}$ without subsampling in any layer, i.e. $s^{(1)}=1$:
\begin{equation}
\textbf{h}_{t}^{(1)}=\textrm{Encoder}^{(1)}(X)\triangleq \textrm{BLSTM}_{t}(X)
\end{equation}
where the BLSTM decoder is labeled as index $1$. 

The combination of CNN and RNN allows the convolutional feature extractor applied on the input to reveal local correlations in both time and frequency dimensions. 
The RNN block on top of CNN makes it easier to learn temporal structure from the CNN output, to avoid modeling direct speech features with more underlying variations.
The pooling layer is essential in CNN to reduce the spatial size of the representation to control over-fitting.
In MEM-Res, we use the initial layers of the VGG net architecture~\cite{simonyan2014very}, stated in table \ref{tab:vgg}, followed by BLSTM layers as VGGBLSTM decoder labeled as index 2:
\begin{equation}
\label{f:vggblstm}
\textbf{h}_{t}^{2}=\textrm{Encoder}^{2}(X)\triangleq \textrm{VGGBLSTM}_{t}(X).
\end{equation}
Two maxpooling layers with $stride=2$ downsample the input features by a factor of $s^{(2)}=4$ in both temporal and spectral directions. 
\begin{table}[th]
  \begin{center}
    	\caption{Initial Six-Layer VGG Configurations}
    \label{tab:vgg}
	\begin{tabular}{cc}
	  \toprule
	  \toprule
{Convolution 2D} & {in = 1, out = 64, filter = 3$\times$ 3}\\
{Convolution 2D} & {in = 64, out = 64, filter = 3$\times$ 3}\\
{Maxpool 2D} & {patch = 2$\times$2, stride = 2$\times$2} \\
{Convolution 2D} & {in = 64, out = 128, filter = 3$\times$ 3}\\
{Convolution 2D} & {in = 128, out = 128, filter = 3$\times$ 3}\\
{Maxpool 2D} & {patch = 2$\times$2, stride = 2$\times$2}\\
	  \bottomrule
	  \bottomrule
	\end{tabular}
  \end{center}
    \vspace{-0.7cm}
\end{table}

\subsection{Multi-Encoder Multi-Array}
\label{sec:proposed}
In this section, we present another realization of multi-stream framework for the multi-array ASR task, i.e. Multi-Encoder Multi-Array (MEM-Array) model. 

\subsubsection{Conventional Multi-Array ASR}
In our previous work, we proposed a stream attention framework to improve the far-field performance in the hybrid approach, using distributed microphone array(s) \cite{wang2018stream}. 
Specifically, we generated more reliable Hidden Markov Model (HMM) state posterior probabilities by linearly combining the posteriors from each array stream, under the supervision of the ASR performance monitors. 

In general, the posterior combination strategy outperformed conventional methods, such as signal-level fusion and the word-level technique ROVER \cite{fiscus1997post}, in the prescribed multi-array configuration. Accordingly, stream attention weights estimated from the de-correlated intermediate features should be more reliable. We adopt this assumption in MEM-Array framework.

\subsubsection{Multi-Array Architecture with Stream Attention}
\label{sssec:muma}
\begin{figure}[htb]
  \centering 
  \vspace{-0.3cm}
  \centerline{\includegraphics[width=7cm]{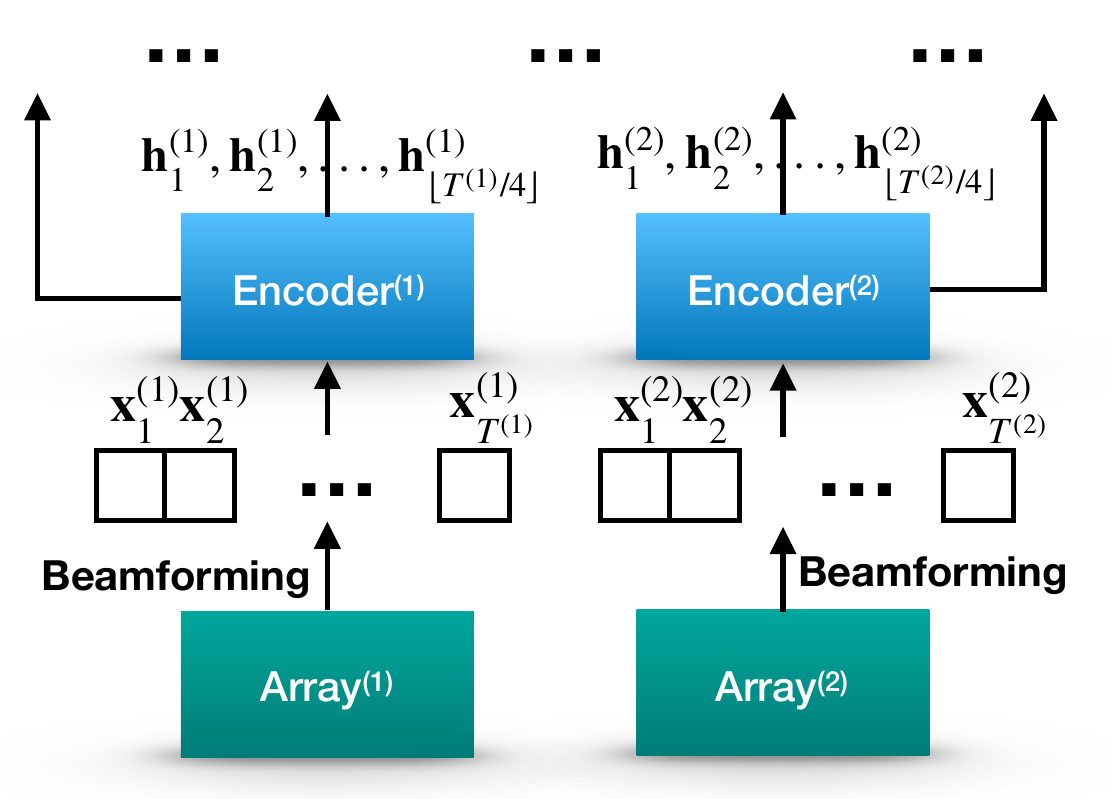}}
\caption{Multi-Encoder Multi-Array Architecture.}
 \vspace{-0.3cm}
  \label{fig:MEM-Array}
\end{figure}
Based on the multi-stream model, the proposed MEM-Array architecture in Fig. \ref{fig:MEM-Array} has two encoders, with each mapping the speech features of a single array to higher level representations ${\bf h}_t^{(i)}$, where we denote $i\in\{1,2\}$ as the index for ${\textrm{Encoder}^{(i)}}$ corresponding to array $i$.
Note that ${\textrm{Encoder}^{(1)}}$ and ${\textrm{Encoder}^{(2)}}$ have the same configurations receiving parallel speech data collected from multiple microphone arrays. 
As we introduced in Sec. \ref{ssec:MEM-Res}, CNN layers are often used together with BLSTM layers on top to extract frame-wise hidden vectors. 
We explore two types of encoder structures: BLSTM (RNN-based) and VGGBLSTM (CNN-RNN-based) \cite{cho2018multilingual}:
\begin{equation}
\textbf{h}_{t}^{(i)}=\textrm{Encoder}^{(i)}(X^{(i)}), i\in\{1,2\}\\
\end{equation}
\begin{equation}
\textrm{Encoder}^{(i)}()=\textrm{BLSTM}() \quad or \quad \textrm{VGGBLSTM}()
\end{equation}

Note that the BLSTM encoders are equipped with an additional projection layer after each BLSTM layer. 
In both encoder architectures, subsampling factor $s^{(1)}=s^{(2)}=4$ is applied to decrease the computational cost. Specially, the convolution layers of the VGGBLSTM encoder downsamples the input features by a factor of 4 so that there is no subsampling in the recurrent layers. 

In the multi-stream setting, one inherent problem is that the contribution of each stream (array) changes dynamically. 
Specially, when one of the streams takes corrupted audio, the network should be able to pay more attention to other streams for the purpose of robustness. 
Inspired by the advances of linear posterior combination \cite{wang2018stream} and a hierarchical attention fusion \cite{yang2016hierarchical, hori2017attention, libovicky2017attention}, a stream-level fusion on the letter-wise context vector is used in this work to achieve the goal of encoder selectivity as we introduced in Sec. \ref{sssec:hieratt}.

In comparison to fusion on frame-wise hidden vectors ${\bf h}_t^{(i)}$, stream-level fusion can deal with temporal misalignment from multiple arrays at the stream level. Furthermore, adding an extra microphone array $j$ could be simply implemented with an additional term $\beta_{l}^{(j)}\textbf{r}_{l}^{(j)}$ in Eq.(\ref{f:han}).

\section{Experiments: MEM-Res Model}
\label{sec:expt}

\subsection{Experimental Setup}
\label{ssec:exptsetup}
We demonstrated our proposed MEM-Res model using two datasets: WSJ1~\cite{wsj1} (81 hours) and CHiME-4~\cite{vincent20164th} (18 hours). 
In WSJ1, we used the standard configuration: ``si284'' for training, ``dev93'' for validation, and ``eval92'' for test. 
The CHiME-4 dataset is a noisy speech corpus recorded or simulated using a tablet equipped with 6 microphones in four noisy environments: a cafe, a street junction, public transport, and a pedestrian area. 
For training, we used both ``tr05\_real'' and ``tr05\_simu'' with additional WSJ1 corpora to support end-to-end training. 
``dt05\_multi\_isolated\_1ch\_track'' was used for validation. 
We evaluated the real recordings with 1, 2, 6-channel in the evaluation set. The BeamformIt \cite{anguera2007acoustic} method was applied to multi-channel evaluation.
In all experiments, 80-dimensional mel-scale filterbank coefficients with additional 3-dimensional pitch features served as the input features.
\begin{table}[ht]
  \begin{center}
   	\caption{Comparison among Single-Encoder End-to-End Models with BLSTM or VGGBSLTM as the Encoder, the MEM-Res Model and Prior End-to-End models. (WER: WSJ1, CHiME-4)}
	\begin{tabular}{lcc}
	  \toprule
	  \toprule
	   &\multicolumn{1}{c}{CHiME-4} & \multicolumn{1}{c}{WSJ1}  \\
 	   Model & et05\_real\_1ch & eval92 \\
	  \midrule
       {\it BLSTM (Single-Encoder)} &  &\\
      CTC   & 62.7 &36.4 \\
      ATT   & 50.2 &20.8 \\
      CTC+ATT  & 29.2 &4.6 \\
      \midrule
      {\it VGGBLSTM (Single-Encoder) } &  \\
      CTC   & 50.6 &19.1 \\
      ATT   & 42.2 &17.2\\
      CTC+ATT   & 29.6 &5.6 \\
      \midrule
      \textcolor{black}{{\it BLSTM+VGGBLSTM (ROVER) }} &  \\
      \textcolor{black}{CTC+ATT} & \textcolor{black}{30.8} & \textcolor{black}{5.9}\\
      \midrule
      {\it BLSTM+VGGBLSTM (MEM-Res) } &  \\
      CTC   & 49.1 &15.2 \\
      ATT   & 44.3 &18.9 \\
      CTC(shared)+ATT   & 26.8 &4.4 \\
      CTC(shared)+ATT+HAN   & 26.9 &4.3 \\
      CTC(per-enc)+ATT  & 26.6 &4.1 \\
      CTC(per-enc)+ATT+HAN   & \textbf{26.4} &\textbf{3.6} \\
      \midrule
      {\it Previous Studies } &  \\
      RNN-CTC \cite{graves2014towards} & - & 8.2\\
      Eesen \cite{miao2015eesen} & - & 7.4 \\
      Temporal LS + Cov. \cite{chorowski2016towards} & - & 6.7 \\
      E2E+regularization\cite{zhou2017improved} & -& 6.3 \\
      Scatt+pre-emp\cite{zeghidour2018end} & -& 5.7 \\
      Joint e2e+look-ahead LM\cite{hori2018end} & -& 5.1 \\
      RCNN+BLSTM+CLDNN \cite{wang2017residual}& -&4.3\\
      EE-LF-MMI \cite{hadian2018end} &-&4.1 \\
      \bottomrule
      \bottomrule
	\end{tabular}
	\label{tab:MEM-Res}
  \end{center}
\end{table}

The $\textrm{Encoder}^{(1)}$ contained four BLSTM layers, in which each layer had 320 cells in both directions followed by a 320-unit linear projection layer. 
The $\textrm{Encoder}^{(2)}$ combined the convolution layers with RNN-based network that had the same architecture as $\textrm{Encoder}^{(1)}$. 
A content-based attention mechanism with 320 attention units was used in encoder-level and frame-level attention mechanisms. 
The decoder was a one-layer unidirectional LSTM with 300 cells. 
We used 50 distinct labels including 26 English letters and other special tokens, i.e., punctuations and sos/eos.

We incorporated the look-ahead word-level RNN-LM~\cite{hori2018end} of 1-layer LSTM with 1000 cells and 65K vocabulary, that is, 65K-dimensional output in Softmax layer. 
In addition to the original speech transcription, the WSJ text data with 37M words from 1.6M sentences was supplied as training data.
RNN-LM was trained separately using Stochastic Gradient Descent (SGD) with learning rate $=0.5$ for 60 epochs. 

The MEM-Res model was implemented using Pytorch backend on ESPnet \cite{watanabe2018espnet}.
Training procedure was operated using the AdaDelta algorithm with gradient clipping on single GPUs, ``GTX 1080ti''. 
The mini-batch size was set to be 15.
We also applied a unigram label smoothing technique to avoid over-confidence predictions.
The beam width was set to 30 for WSJ1 and 20 for CHiME-4 in decoding. 
For model jointly trained with CTC and attention objectives, $\lambda=0.2$ was used for training, and $\lambda=0.3$ for decoding.
RNN-LM scaling factor $\gamma$ was $1.0$  for all experiments with the exception of using $\gamma=0.1$ in decoding attention-only models.

\subsection{Results}
\label{ssec:results}

The overall experimental results on WSJ1 and CHiME-4 are shown in Table \ref{tab:MEM-Res}.
Compared to joint CTC/Attention single-encoder models, the proposed MEM-Res model with per-encoder CTC and HAN achieved relative improvements of \textcolor{black}{$9.6\%$ ($29.2\%\rightarrow 26.4\%$)} in CHiME-4 and 21.7\% in WSJ1 ($4.6\%\rightarrow 3.6\%$) in terms of WER. 
We compared the MEM-Res model with other end-to-end approaches, and it outperformed all of the systems from previous studies. 
\textcolor{black}{We also conducted experiments using ROVER technique \cite{rover} to fuse two single-encoder models in the word level, and our proposed models showed substantial improvements. }
We designed experiments with fixed encoder-level attention ${\beta}_{l}^{1}={\beta}_{l}^{(2)}=0.5$. And the MEM-Res model with HAN outperformed the ones without parameterized stream attention.
Moreover, per-encoder CTC constantly enhanced the performance with or without HAN. Specially in WSJ1, the model shows notable decrease ($4.3\%\rightarrow 3.6\%$) in WER with per-encoder CTC. 
Our results further confirmed the effectiveness of joint CTC/Attention architecture in comparison to models with either CTC or attention network. 

\begin{table}[!htbp]
  \begin{center}
   	\caption{Comparison between the MEM-Res Model and VGGBSLTM Single-Encoder Model with Similar Network Size. (WER: WSJ1, CHiME-4)}
	\begin{tabular}{lcc}
	  \toprule
	  \toprule
	  & Single-Encoder & Proposed Model\\
	  Data & (21.9M) & (21.3M) \\
      \midrule
      {\it CHiME-4} & \\
      et05\_real\_1ch & 32.2 & \textbf{26.4 (18.0\%)} \\
      et05\_real\_2ch & 26.8 & \textbf{21.9 (18.3\%)} \\
      et05\_real\_6ch & 21.7 & \textbf{17.2 (20.8\%)} \\
      \midrule
	  {\it WSJ1}  & \\
      eval92 & 5.3 & \textbf{3.6 (32.1\%)} \\
      \bottomrule
      \bottomrule
	\end{tabular}
	\label{tab:cmp1stream}
  \end{center}
     \vspace{-0.1cm}
\end{table}
For fair comparison, we increased the number of BLSTM layers from 4 to 8 in $\textrm{Encoder}^{(2)}$ to train a single-encoder model. 
In Table \ref{tab:cmp1stream}, the MEM-Res system outperforms the single-encoder model by a significant margin with similar amount of parameters, $21.9$M v.s. $21.3$M.
In CHiME-4, we evaluated the model using real test data from 1, 2, 6-channel resulting in an average of $\textbf{19\%}$ relative improvement from all three setups.
In WSJ1, we achieved $\textbf{3.6\%}$ WER in eval92 in our MEM-Res framework with relatively $\textbf{32.1\%}$ improvement.


\begin{table}[!htbp]
  \begin{center}
   	\caption{Effect of Multi-Resolution Configuration $(s^{(1)},s^{(2)})$, where  $s^{(1)}$ and $s^{(2)}$ are Subsampling Factors for $\textrm{Encoder}^{(1)}$ and $\textrm{Encoder}^{(2)}$. (WER: WSJ1, CHiME-4)}
	\begin{tabular}{lccc}
	  \toprule
	  \toprule
	  Data & (4,4) & (2,4) & (1,4) \\
	  \midrule
      {\it CHiME-4} & & & \\	  
      et05\_real\_1ch & 29.1& 27.0 &\textbf{26.4}\\
      \midrule
      {\it WSJ1}  & & \\
      eval92 &  4.5& 4.2&\textbf{3.6}\\
      \bottomrule
      \bottomrule
	\end{tabular}
    \label{tab:ss}
  \end{center}
    \vspace{-0.3cm}
\end{table}
The results in Table \ref{tab:ss} shows the contribution of multiple resolution.  
The WER went up when increasing subsampling factor $s^{(1)}$ closer to $s^{(2)}=4$ in both datasets. In other words, the fusion worked better when two encoders are more heterogeneous which supports our hypothesis. 
As shown in Table \ref{tab:att}, We analyzed the average stream-level attention weight for $\textrm{Encoder}^{(2)}$ when we gradually decreased the number of LSTM layers while keeping $\textrm{Encoder}^{(1)}$ with the original configuration. It aimed to show that HAN was able to attend to the appropriate encoder seeking for the right knowledge. As suggested in the table, more attention goes to $\textrm{Encoder}^{(1)}$ from $\textrm{Encoder}^{(2)}$ as we intentionally make $\textrm{Encoder}^{(2)}$ weaker.
\begin{table}[!htbp]
  \begin{center}
   	\caption{Analysis of Hierarchical Attention Mechanism when Fixing $\textrm{Encoder}^{(1)}$ and Changing the Number of LSTM Layers in $\textrm{Encoder}^{(2)}$. (WER: CHiME-4)}
	\begin{tabular}{c|cc}
	  \toprule
	  \toprule
      \# LSTM Layers &\multicolumn{1}{c}{Average Stream Attention} &   \\
	  in VGGBLSTM & for VGGBLSTM & WER \% \\
	  \midrule
	   0& 0.27 & 30.6\\
       1& 0.52 & 29.8\\
       2 & 0.75 & 28.9\\
	   3 & 0.82 & 27.8\\
       4 & 0.81 & 26.4\\
      \bottomrule
      \bottomrule
	\end{tabular}
	\label{tab:att}
  \end{center}
    \vspace{-0.7cm}
\end{table}

\section{Experiments: MEM-Array Model}
\label{sec:experiment}

\subsection{Experimental Setup}
\label{amidescrip}
\textcolor{black}{Two dataset, AMI Meeting Corpus and DIRHA English WSJ, were used to demonstrate MEM-Array model.
The AMI meeting corpus \cite{carletta2005ami} was created in three instrumented meeting rooms focusing on developing meeting browsing technology.
There are 100 hours of far-field signal-synchronized recordings collected using two microphone arrays placed in each room.
The training, development and evaluation set are comprised of 81 hours, 9 hours and 9 hours of meeting recordings, respectively.
The DIRHA English WSJ \cite{ravanelli2016realistic} was part of DIRHA project which addresses the challenge of speech interaction via distant microphones.
A total of 32 microphones were used in a domestic environment of a living room and a kitchen.
Two microphone arrays, a circular array and a linear array in the living room, were chosen as parallel streams.
Contaminated version of the original WSJ0 and WSJ1 corpus was used for training, providing room impulse responses for correspoding arrays.
Development set for cross validation was simulated with typical domestic background noise and reverberation.
Evaluation set has 409 read utterances from WSJ text recorded by six native English speakers in real domestic setting.}

\textcolor{black}{For both datasets, two microphone arrays (noted by Str1 and Str2) were applied to train a MEM-Array model, where configuration of arrays for each dataset is described in Table \ref{tab:table1}.
Note that for each array, multi-channel input was synthesized into a single-channel audio using Delay-and-Sum beamforming technique with BeamformIt Toolkit \cite{anguera2007acoustic}.
Experiments were conducted with configuration as described in Table \ref{tab:config}.}
\begin{table}[htb]
  \begin{center}
   	\caption{Description of the Array Configuration in the Two-Stream E2E Experiments.}
    \label{tab:table1}
	\begin{tabular}{l|c|c}
	  \toprule
	  Dataset   & Str1 (Stream 1)  & Str2 (Stream 2) \\
	  \midrule
             &                      & Edinburgh: 8-mic Circular Array \\
      AMI    & 8-mic Circular Array & Idiap: 4-mic Circular Array \\
             &                      & TNO: 10-mic Linear Array \\
      \midrule
      DIRHA  & 6-mic Circular Array & 11-mic Linear Array \\
      \bottomrule
	\end{tabular}
  \end{center}
\end{table}
\begin{table}[th]
  \begin{center}
   	\caption{Experimental Configuration (MEM-Array)}
    \label{tab:config}
	\begin{tabular}{ll}
	  \toprule
	  {\scriptsize{}{\bf Feature}} \\
      {\scriptsize{}Single Stream} & {\scriptsize{}80-dim fbank + 3-dim pitch}\\
      {\scriptsize{}Multi Stream} & {\scriptsize{}$\text{Array}^{(1)}$:80+3; $\text{Array}^{(2)}$:80+3}\\
      \hline
	  {\scriptsize{}{\bf Model}} \\
      {\scriptsize{}Encoder type} & {\scriptsize{}BLSTM or VGGBLSTM}\\
      {\scriptsize{}Encoder layers} & {\scriptsize{}BLSTM:4; VGGBLSTM\cite{cho2018multilingual}:6(CNN)+4}\\
      {\scriptsize{}Encoder units } & {\scriptsize{}320 cells (BLSTM layers)}\\
      {\scriptsize{}(Stream) Attention} & {\scriptsize{}Content-based}\\
      {\scriptsize{}Decoder type} & {\scriptsize{}1-layer 300-cell LSTM}\\
      {\scriptsize{}CTC weight $\lambda$ (train)} & {\scriptsize{}AMI:0.5; DIRHA:0.2}\\
    {\scriptsize{}CTC weight $\lambda$ (decode)} & {\scriptsize{}AMI:0.3; DIRHA:0.3}\\
        \hline 
        {\scriptsize{}{\bf RNN-LM}} \\
      {\scriptsize{}Type} & {\scriptsize{}Look-ahead Word-level RNNLM \cite{hori2018end}}\\
      {\scriptsize{}Train data} & {\scriptsize{}AMI:AMI; DIRHA:WSJ0-1+extra WSJ text data}\\
      {\scriptsize{}LM weight $\gamma$} & {\scriptsize{}AMI:0.5; DIRHA:1.0}\\
      \bottomrule
	\end{tabular}
  \end{center}
    \vspace{-0.3cm}
\end{table}
\subsection{Results}
\textcolor{black}{Similar to experiments in MEM-Res session, we started with discussion on single-stream architecture, followed by analysis of the effectiveness of our proposed MEM-Array model.}

\textcolor{black}{Results for single-array models are summarized in Table \ref{tab:table3}. By comparing two encoder architectures on both datasets, VGGBLSTM noticeably outperforms BLSTM as encoder type. 
With the help of CTC and an external RNNLM, substantial improvements were observed throughout all the cases of Stream 1. 
The architecture with the best performance (VGGBLSTM+CTC+ATT+RNNLM) was chosen for further experiments on Stream 2 in Table \ref{tab:table3}.}
\begin{table}[htb]
  \begin{center}
  	\caption{Exploration of Best Encoder and Decoding Strategy for Single-Stream E2E Model.}
    \label{tab:table3}
	\begin{tabular}{l|c|c|c|c}
	  \toprule
	  &\multicolumn{2}{|c|}{AMI} & \multicolumn{2}{c}{DIRHA}  \\
	  Model (Single Stream) & \multicolumn{2}{c|}{Eval}0 & \multicolumn{2}{c}{Real}  \\
	        & CER & WER &  CER & WER \\
	  \midrule
      {\it BLSTM} (Str1)\ \ \  & & & \\
      Attention  & 45.1 & 60.9 & 42.7 & 68.7 \\
      + CTC      & 41.7	& 63.0 & 38.5 & 74.8\\
      + Word RNNLM   & 41.7	& 59.1 & 29.4 & 47.4\\
      \midrule
      {\it VGGBLSTM} (Str1)\ \ \  & & & \\
      Attention  & 43.2	& 59.7 & 39.5 &	71.4 \\
      + CTC      & 40.2	& 62.0 & 30.1 & 61.8\\
      + Word RNNLM   & \bf 39.6	& \bf 56.9 & \bf 21.2 & \bf 35.1\\
      \midrule
      \midrule
      {\it VGGBLSTM} (Str2)\ \ \  & \bf 45.6	& \bf 64.0 & \bf 22.5 & \bf 38.4 \\
      \bottomrule
	\end{tabular}
  \end{center}
  \vspace{-0.0cm}
\end{table}

\textcolor{black}{As illustrated in Table \ref{tab:table4}, our proposed framework was able to fuse information successfully from both streams by achieving lower error rates than best single-array systems, i.e., AMI ($56.9\%\rightarrow 54.9\%$) and DIRHA ($35.1\%\rightarrow 31.7\%$).
Moreover, several conventional fusion strategies were discussed in Table \ref{tab:table4}: 
signal-level fusion through WAV alignment and average; feature-level frame-by-frame concatenation; word-level prediction fusion using ROVER. 
The MEM-Array model outperformed all three fusion techqinues, even including the case when doubling BLSTM layers in singal-level fusion for a comparable amount of parameters (33.7M vs 31.6M).}
\begin{table}[htb]
  \begin{center}
   	\caption{WER(\%) Comparison between Proposed Multi-Stream Approach and Alternative Single-Stream Strategies.}
    \label{tab:table4}
	\begin{tabular}{l|c|c|c}
	  \toprule
	  Encoder {\it VGGBLSTM}  & \#Param & AMI & DIRHA \\
	  (Att + CTC + RNNLM)      &    & Eval  & Real \\
	  \midrule
	  {\it Single-stream model} & &\\
	  Concatenating Str1\&Str2 & 23.3M & 56.7 & 33.5 \\
	  WAV alignment and average & 26.2M & 56.7 & 43.5 \\
          + model parameter extension  & 33.7M & 56.9 & 39.6 \\
        \textcolor{black}{{\it Two single-stream models} }& &\\
        	  \textcolor{black}{ROVER Str1\&Str2} &\textcolor{black}{ 52.5M(26.2$\times$2)} & \textcolor{black}{60.7} & \textcolor{black}{37.0}\\
	  \midrule
	  \midrule
      {\it Multi-stream model}  & &   \\
      Proposed framework & 31.6M &  \bf 54.9 & \bf 31.7 \\
      \bottomrule
	\end{tabular}
  \end{center}
    \vspace{-0.2cm}
\end{table}

\begin{figure}[tb]
  \centering 
  \centerline{\includegraphics[width=8.5cm]{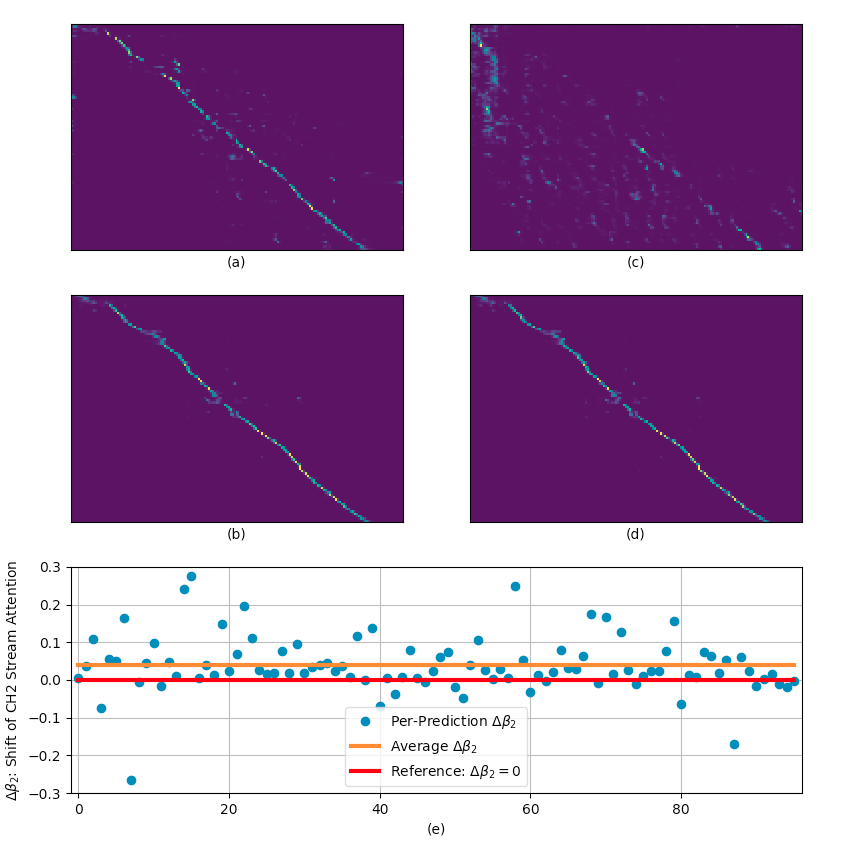}}
\caption{Comparison of the alignments between characters (y-axix) and acoustic frames (x-axis) before (({\bf a}) Str1; ({\bf b}) Str2) and after (({\bf c}) Str1; ({\bf d}) Str2) noise corruption of Str1. ({\bf e}) shows the attention weight shift of Str2 between two cases (x-axis is the letter sequence).}
  \label{fig:corr}
  \vspace{-0.3cm}
\end{figure}

\textcolor{black}{To investigate robustness of stream attention, we designed an experiment with Str1 injected with zero-mean, unit-variance Gaussian noise in the signal level while keeping Str2 untouched.
Fig.\ref{fig:corr} displays an example from DIRHA evaluation set during inference. 
Noise corruption Str1 ($(a)\rightarrow (c)$) made attention alignments fairly blurred, thus less trusted. 
As expected, an averagely positive shift of stream attention weights towards Str2 was observed.}
\textcolor{black}{Table \ref{tab:table5} shows fusion results of six streams in hybrd ASR system from our previous study \cite{wang2018stream}. 
Relative WER reductions of 7.2\% and 5.8\% were reported comparing to the best single stream performance. 
Meanwhile, MEM-Array system with two streams reduced the WER by 9.7\% relatively. 
In spite of more training data involved in E2E, MEM-Array shows a promising direction for fusion of more streams.}

\begin{table}[htb]
  \begin{center}
   	\caption{WER(s) Comparison between the Hybrid and End-to-End System on DIRHA Dataset. \#Num Denotes the Number of Streams.}
    \label{tab:table5}
	\begin{tabular}{l|c|c|c|c}
	  \toprule
	  System & \#Num & Method & Best Stream & WER  \\
	  \midrule
	  Hybrid   & 6 & post. comb. & 29.2 & 27.1 (\bf 7.2\%) \\
	           & 6 & ROVER & 29.2 & 27.5 (\bf 5.8\%)\\
	  \midrule
	  E2E & 2 & proposed & 35.1 & 31.7 (\bf 9.7\%)\\
      \bottomrule
	\end{tabular}
  \end{center}
\end{table}

\section{conclusion}
\label{sec:conclusion}
In this work, we present our multi-stream framework to build an end-to-end ASR system. 
Higher-level frame-wise acoustic features were carried out from parallel encoders with various configurations of input features, architectures and temporal resolutions. 
Stream attention was achieved through a hierarchical connection between the decoder and encoders.
We also investigated that assigning a CTC network to individual encoder further  helped  diverse  encoders to reveal complementary information.

Two realizations of multi-stream framework have been proposed, which are MEM-Res model and MEM-Array model targeting different applications. 
In MEM-Res architecure, RNN-based and CNN-RNN-based encoders with subsampling only in convolutional layers characterized same speech in different ways.
The model outperformed various single-encoder models, reaching the state-of-the-art performance on WSJ among end-to-end systems. 
\textcolor{black}{For further study, exploring hierarchical feedback from different decoder layers  and advanced convolutional layers, such ResNet, and self-attention layers have the potential to improve the WER even more.} 
In multi-array scenarios, taking advantage of all the information that each array shared and contributed was crucial in this task. 
The MEM-Array model represent each array with one encoder followed by attention fusion in the contextual vector level, where no frame synchronization of parallel stream was required. 
Thanks to the success of joint training of per-encoder CTC and attention, substantial WER reduction was shown in both AMI and DIRHA corpora, demonstrating the potentials of the proposed architecture. 
An extension to more streams efficiently and exploration of schedule training of the encoders are to be investigated.

\appendices


\section*{Acknowledgment}

This work is supported by National Science Foundation under Grant No. 1704170 and No. 1743616, and a Google faculty award to Hynek Hermansky.

\ifCLASSOPTIONcaptionsoff
  \newpage
\fi



\bibliographystyle{IEEEtran}
\bibliography{IEEEabrv,refs}
%




%

\begin{IEEEbiography}[{\includegraphics[width=1in,height=1.25in,clip,keepaspectratio]{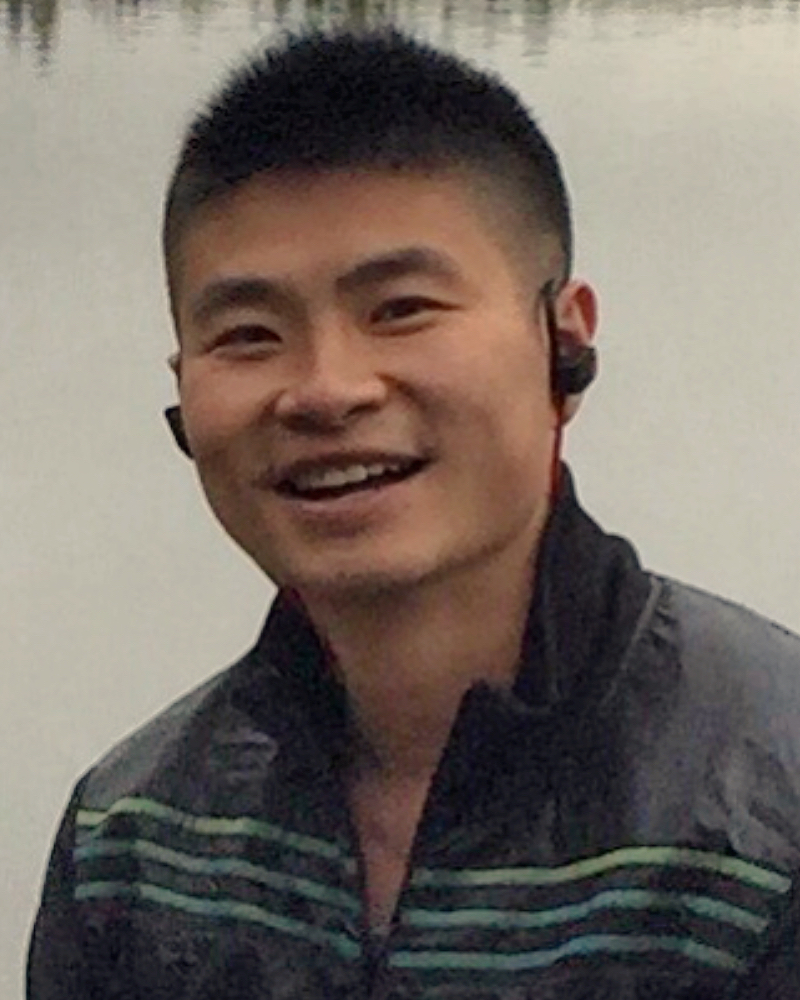}}]{Ruizhi Li}
is a Ph.D. student at Johns Hopkins University since 2014. His research interests include machine learning and spoken language processing. He received his B.E. degree in Electrical Engineering in Beijing University of Chemical Technology in 2012, and M.S. degree in Electrical Engineering from Washington University in St. Louis in 2014. He is a student member of the IEEE.
\end{IEEEbiography}

\begin{IEEEbiography}[{\includegraphics[width=1in,height=1.25in,clip,keepaspectratio]{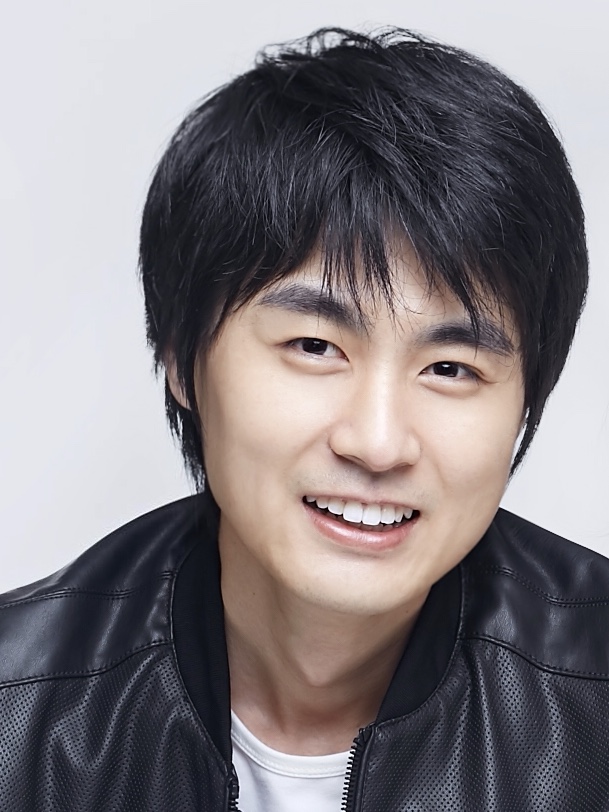}}]{Xiaofei Wang}
is a postdoctoral research fellow of Center for Language and Speech Processing at Johns Hopkins University in Baltimore, MD, USA, since 2016. He received the Ph.D. from University of Chinese Academy of Sciences in 2015 and B.E. from Huazhong University of Science and Technology, China in 2010. From 2015 to 2016, he was an Assistant Professor at Institute of Acoustics, Chinese Academy of Sciences.
His research interests are far-field automatic speech recognition and speech enhancement. He is member of IEEE and ISCA.
\end{IEEEbiography}

\begin{IEEEbiography}[{\includegraphics[width=1in,height=1.25in,clip,keepaspectratio]{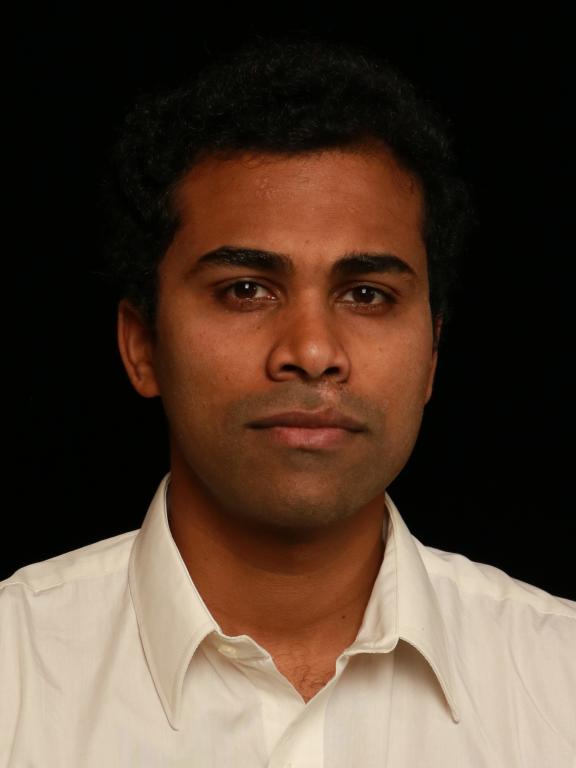}}]{Sri Harish Mallidi} is an applied scientist in Amazon, Seattle, USA, where he is working on algorithms and technologies for large-scale, real-time automatic speech recognition systems. He received his Doctor of Philosophy from the Center for Language and Speech Processing, Johns Hopkins University in 2018 with Prof. Hynek Hermansky. Prior to this, he obtained his B.Tech (2008) and M.S. (2010) in Electronics and Communications from International Institute of Information Technology, Hyderabad (IIIT-H), India.  His research interests include machine learning methods for speech recognition, speech activity detection, keyword spotting, and speaker recognition and diarization. 
\end{IEEEbiography}

\begin{IEEEbiography}[{\includegraphics[width=1in,height=1.25in,clip,keepaspectratio]{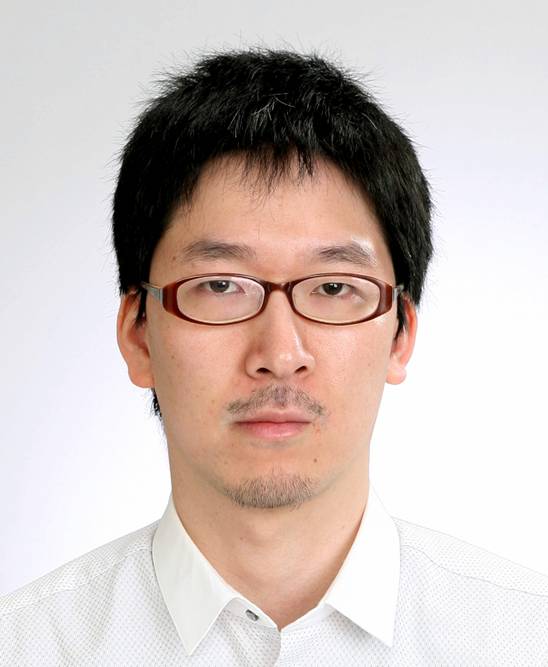}}]{Shinji Watanabe}
is an Associate Research Professor at Johns Hopkins University, Baltimore, MD, USA. 
He received his B.S., M.S. PhD (Dr. Eng.) Degrees in 1999, 2001, and 2006, from Waseda University, Tokyo, Japan. 
He was a research scientist at NTT Communication Science Laboratories, Kyoto, Japan, from 2001 to 2011, a visiting scholar in Georgia institute of technology, Atlanta, GA in 2009, and a Senior Principal Research Scientist at Mitsubishi Electric Research Laboratories (MERL), Cambridge, MA from 2012 to 2017. 
His research interests include automatic speech recognition, speech enhancement, spoken language understand, and machine learning for speech and language processing. He has been published more than 150 papers in top journals and conferences, and received several awards including the best paper award from the IEICE in 2003. 
He served an Associate Editor of the IEEE Transactions on Audio Speech and Language Processing, and is a member of several technical committees including the IEEE Signal Processing Society Speech and Language Technical Committee (SLTC) and Machine Learning for Signal Processing Technical Committee (MLSP).
\end{IEEEbiography}

\begin{IEEEbiography}[{\includegraphics[width=1in,height=1.25in,clip,keepaspectratio]{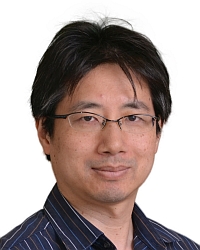}}]{Takaaki Hori}
(SM'14) received the B.E. and M.E. degrees in electrical and information engineering from Yamagata University, Yonezawa, Japan, in 1994 and 1996, respectively, and the Ph.D. degree in system and information engineering from Yamagata University in 1999.
From 1999 to 2015, he had been engaged in researches on speech recognition and spoken language understanding at Cyber Space Laboratories and Communication Science Laboratories in Nippon Telegraph and Telephone (NTT) Corporation, Japan. He was a visiting scientist at the Massachusetts Institute of Technology (MIT) from 2006 to 2007. Since 2015, he has been a senior principal research scientist at Mitsubishi Electric Research Laboratories (MERL), Cambridge, Massachusetts, USA. 
He has coauthored more than 100 peer-reviewed papers in speech and language research fields.
He received the 24th TELECOM System Technology Award from the Telecommunications Advancement Foundation in 2009, the IPSJ Kiyasu Special Industrial Achievement Award from the Information Processing Society of Japan in 2012, and the 58th Maejima Hisoka Award from Tsushinbunka Association in 2013.
\end{IEEEbiography}

\begin{IEEEbiography}[{\includegraphics[width=1in,height=1.25in,clip,keepaspectratio]{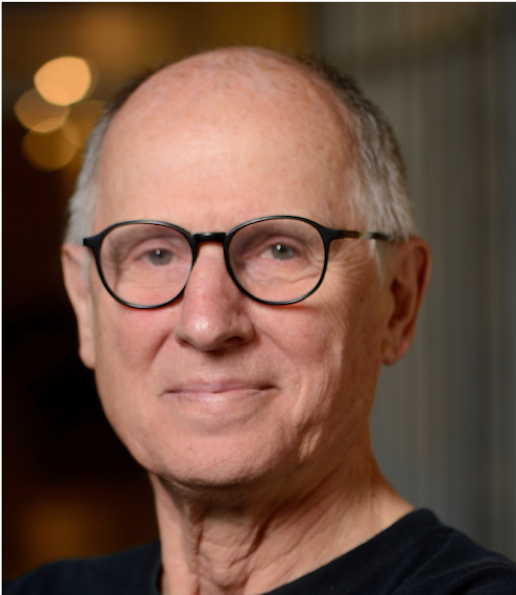}}]{Hynek Hermansky}
(LF’17, F'01, SM'92. M'83, SM'78) received the Dr. Eng. Degree from the University of Tokyo, and Dipl. Ing. Degree from Brno University of Technology, Czech Republic. 

He is the Julian S. Smith Professor of Electrical Engineering and the Director of the Center for Language and Speech Processing at the Johns Hopkins University in Baltimore, Maryland.  He is also a Professor at the Brno University of Technology, Czech Republic. He has been working in speech processing for over 30 years.
His main research interests are in acoustic processing for speech recognition.

He is a Life Fellow of IEEE, and a Fellow of the International Speech Communication Association (ISCA), He is the General Chair of the INTERSPECH 2021, was the General Chair of the 2013 IEEE Automatic Speech Recognition and Understanding Workshop, was in charge of plenary sessions at the 2011 ICASSP in Prague, was the Technical Chair at the 1998 ICASSP in Seattle and an Associate Editor for IEEE Transaction on Speech and Audio. He is also a Member of the Editorial Board of Speech Communication, was twice an elected Member of the Board of ISCA, a Distinguished Lecturer for IEEE, a Distinguished Lecturer for ISCA, and the recipient of the 2013 ISCA Medal for Scientific  Achievement.
\end{IEEEbiography}







\end{document}